\documentclass[epj,final]{svjour}  
\usepackage{graphics}

\begin{document}
\sloppy

\title{
Potential and limitations of nucleon transfer experiments with radioactive
beams at REX-ISOLDE}
\author{%
C.~Gund\inst{1}\thanks{\emph{Present address:} Fraunhofer IIS-B, D-91058 Erlangen, Germany} \and 
H.~Bauer\inst{1}\thanks{\emph{Present address:} McKinsey \& Company, D-80538 M\"unchen, Germany} \and 
J.~Cub\inst{2}\thanks{\emph{Present address:}  DePfa Systems GmbH, D-55122 Mainz, Germany} \and 
A.~Dietrich\inst{1}\thanks{\emph{Present address:} Fachbereich Materialwissenschaft, Technische Universit\"at Darmstadt, D-64287 Darmstadt, Germany} \and
T. H\"artlein\inst{1}\thanks{\emph{Present address:} SAP AG, 69190 Walldorf, Germany} \and 
H. Lenske\inst{3}     \and
D.~Pansegrau\inst{1}\thanks{\emph{Present address:} Robert Bosch GmbH, 70442 Stuttgart, Germany} \and
A.~Richter\inst{2}    \and
H.~Scheit\inst{1}     \and
G.~Schrieder\inst{2}  \and
D. Schwalm\inst{1}
} 
\titlerunning{Transfer experiments at REX-ISOLDE}
\authorrunning{C. Gund \textit{et al.}}
\offprints{heiko.scheit@mpi-hd.mpg.de}
\mail{heiko.scheit@mpi-hd.mpg.de}
\institute{
Max-Planck-Institut f\"ur Kernphysik, D-69117 Heidelberg, Germany, \email{initial.name@mpi-hd.mpg.de} \and
Institut f\"ur Kernphysik, Technische Universit\"at Darmstadt, D-64298 Darmstadt, Germany \and
Institut f\"ur Theoretische Physik, Universit\"at Gie\ss en, D-35392 Gie\ss en, Germany
} 
\date{Received: / Accepted:}

\abstract{
As a tool for studying the structure of nuclei far off stability the technique of
$\gamma$-ray spectroscopy after low-energy
single-nucleon transfer reactions with radioactive nuclear beams in inverse kinematics
was investigated. Modules of the MINIBALL germanium array and a thin
position-sensitive parallel plate avalanche counter (PPAC) to be employed in
future experiments at REX-ISOLDE were used in a test experiment performed with a stable 
$^{36}$S beam on deuteron and $^9$Be targets. It is demonstrated that the 
Doppler broadening of $\gamma$ lines detected by the MINIBALL modules is considerably 
reduced by exploiting their segmentation, and that for beam intensities up to 10$^6$
particles/s the PPAC positioned around zero degrees with respect to
the beam axis allows not only to significantly reduce the 
$\gamma$ background by requiring coincidences with the transfer products but also to control
the beam and its intensity by single particle counting.
The predicted large neutron pickup cross sections of neutron-rich light nuclei 
on $^2$H and $^9$Be targets at REX-ISOLDE energies of 2.2~MeV$\cdot A$ are confirmed. 
} 
\maketitle
\section{Introduction} 
REX-ISOLDE \cite{habs}, an experiment presently constructed at the ISOLDE facility at CERN, is scheduled
to supply beams of exotic nuclei with energies of up to 2.2~MeV$\cdot A$ starting in 
2001. At these energies Coulomb
excitation and transfer reactions in inverse kinematics on light target materials can be used
to investigate the collective and single-particle structure of low-lying levels of nuclei with extreme
$N/Z$-ratios.

REX-ISOLDE uses the 60~keV 1$^+$~beams delivered by the ISOLDE facility and 
accelerates them to 2.2~MeV$\cdot A$ using a novel acceleration scheme employing
a buncher-trap, a charge-breeder, an intermediate mass-selector and three
RF accelerator units. While for isotopes close to stability beam
intensities of up to $I=10^{10}$ particles/s
can be expected, the intensities decline rapidly  
as one approaches the neutron or proton dripline
(e.g. ${\rm I}(^{24}{\rm Na})\approx 5\cdot 10^{8}~{\rm s}^{-1}$,
${\rm I}(^{31}{\rm Na})\approx 50~{\rm s}^{-1}$).

These low beam intensities require experimental setups
optimized to achieve both high reaction rates and high detection efficiencies
without compromising resolution.
Since for particle spectroscopy thin
targets and high resolution spectrometers are needed, which usually have only
small angular acceptances, counting rates will be low in such experiments.
In contrast, $\gamma$-ray spectroscopy with a highly efficient
germanium detector array offers the possibility to use
thicker targets while maintaining excellent resolution. 
However, in experiments at low beam intensities the background
$\gamma$ rays usually outnumber the
nuclear reaction $\gamma$ rays by orders of magnitude, 
a problem which is even more pronounced when radioactive beams are
used due to the $\beta$-decay
background.
It is  therefore advantageous to detect at least one charged particle produced in the reaction in a
large solid angle and to require
a time coincidence with the $\gamma$ ray to
achieve a sufficient background suppression. 
The most favorable approach for inverse reactions on very light target materials like $^2$H and at low beam
intensities is to detect the heavy reaction product, whose momentum hardly differs
from the momentum of the beam particle.
Thus the heavy reaction products can easily be detected
within their full solid angle by a detector positioned around
zero degrees with respect to the beam axis.  
Alternatively, the detection of the beam
particle in front of the target is possible, yet this would cause an
extra energy loss of the beam before it reaches the target.

$\gamma$-ray spectroscopy at low beam intensities
requires a germanium array with excellent photopeak efficiency. 
As in REX-ISOLDE mainly low-lying levels of exotic nuclei will be
investigated, the $\gamma$ cascades are expected to be of low
multiplicity. 
It is therefore possible to achieve the high efficiency by
placing the germanium detectors very close to the target, which results in
a considerable cost reduction per percent efficiency as only a 
moderate number of germanium modules is required. 
Yet, the close geometry has an important disadvantage. 
The large recoil
velocities cause a considerable Doppler broadening of the $\gamma$ lines as the width of
a Doppler broadened peak increases with the solid angle covered by a
single detector module. 
Internal segmentation of the modules is the
best remedy for this problem as this does not sacrifice photopeak efficiency.

MINIBALL \cite{eberth}, a germanium array which was especially developed for
low-multiplicity, high-efficiency $\gamma$-ray spectroscopy experiments
will thus be used at REX-ISOLDE. 
Extrapolating from the measured
efficiencies
for a single module, MINIBALL will have a 4$\pi$ efficiency of about
15.5\% at 646~keV and 12.5\% at 1333~keV.
For the individual MINIBALL modules the encapsulation technique
and the semi-hexaconical shape of the EUROBALL modules \cite{thomas}
were employed, and with a length of 78~mm and a front diameter of 68~mm they are also of equal size. 
However, the MINIBALL modules are electrically segmented into six parts 
(see inset in figure~\ref{bild.aufbau}), which results in a six-fold enhanced
granularity. 
In the experiment described below, standard
electronics for the energy readout of the core and the six
segments were used and only for the current signal of the core a
pulse shape analysis was possible.
New
electronics for the MINIBALL array are presently commissioned, which
will allow pulse shape analysis of the charge signals of both the core 
and the six segments while maintaining excellent energy resolution. 
With the aid
of these electronics and advanced algorithms \cite{gund,weisshaar} 
to analyze the charge signals an improvement of the module granularity 
of almost one order of magnitude \cite{gund} can be achieved 
as compared to that of the six-fold segmented module without pulse shape 
analysis.

In order to show the feasibility of $\gamma$-ray
spectroscopy experiments using inverse transfer
reactions with low-intensity beams
a test experiment was performed at the 
tandem accelerator facility of the 
Max-Planck-Institut f\"ur Kernphysik, Heidelberg.
In the measurement $^{36}$S beams at the maximum REX-ISOLDE energy of
2.2~MeV$\cdot A$ and with intensities as low as
2$\cdot$10$^5$ particles/s were used together with targets of deuterium enriched polythene 
and $^9$Be. For the deuterium target both
proton and neutron pickup reactions were expected to take place, whereas only neutron
pickup is expected using a beryllium target, as here the proton transfer
is kinematically suppressed.
 
One aim of the experiment
was to verify the predicted high transfer cross sections for
single neutron transfer at REX-ISOLDE beam energies \cite{lenske}.
The second aim was to determine the $\gamma$-energy resolution after Doppler correction for the
MINIBALL modules under realistic conditions. 
Furthermore, it was aimed to examine how the few reaction $\gamma$ rays
expected in low beam-intensity experiments can be effectively separated from the background.
For this purpose the new REX-ISOLDE Parallel
Plate Avalanche Counter (PPAC) \cite{cub} was mounted under zero degrees with
respect to the beam to detect the heavy
transfer products and to produce a reference signal
for particle-$\gamma$-coincidence measurements.

\section{Inverse Reaction Kinematics and Differential Cross Sections}
\label{sektion.kinematik}
In this section the test reaction of $^{36}$S on $^2$H and $^9$Be at 
a beam energy of 2.2 MeV$\cdot A$ is used to exemplify the kinematics 
of a typical inverse transfer reaction to be studied at REX-ISOLDE.

Since neutrons are not available as a target material, stable nuclei which 
have a weakly bound neutron should be used in neutron pickup experiments.
Mainly two light target materials are available, which permit neutron-transfer
reactions in inverse reaction kinematics at REX-ISOLDE energies:
deuterium and beryllium. 
As the deuteron consists of a proton and a neutron bound with an energy of
2.2~MeV, besides the neutron pickup channel $^2$H($^{36}$S,$^{37}$S)p,
the proton pickup channel $^2$H($^{36}$S,$^{37}$Cl)n will occur
when bombarding a $^2$H target with 2.2~MeV$\cdot A$ $^{36}$S.
Other reaction channels are possible, but can be neglected at small beam energies.
The two main channels cannot be distinguished unless the Q-values are very different
(as is the case for exotic beams, but not for the present experiment), the proton 
is detected, or the level scheme of one of the two transfer products is already known. 
With a neutron separation energy of 1.67~MeV and a proton separation energy of
16.9~MeV $^9$Be can also be considered to be a perfect neutron target. 
In contrast to deuterium, where in the present case the proton and neutron pickup
can occur with equal cross sections,
the proton pickup is strongly suppressed for the beryllium target. 
Yet, there exist several other
reaction channels, where many nucleons are transferred from
$^9$Be to the projectile (see section~\ref{kap.beryllium}).

\begin{figure}
\begin{center}
\resizebox{0.5\textwidth}{!}{%
\includegraphics{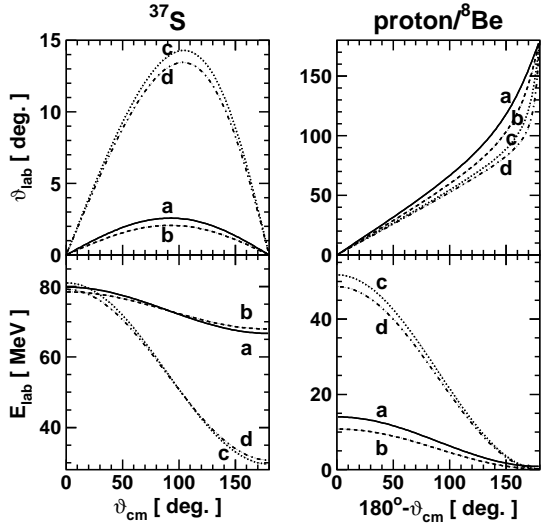}}
\end{center}
\caption[Kinematics of the Pickup Reactions]
{Kinematics of the $^2$H($^{36}$S,$^{37}$S$^*$)p (a,b) and
  $^9$Be($^{36}$S,$^{37}$S$^*$)$^8$Be (c,d) reaction populating the
  3/2$^-$ level at 646~keV (a,c) and the 1/2$^-$ level in
  $^{37}$S at 2.64~MeV (b,d). 
  Here $\vartheta_\mathrm{cm}$ is the scattering angle of $^{37}$S
  in the c.m. system, while $\vartheta_\mathrm{lab}$ denotes the
  laboratory angle and $E_\mathrm{lab}$ the laboratory energy of
  one of the two reaction products. 
  All angles are measured with respect to the beam
  axis. 
\label{bild.kinematik}
}
\end{figure}

The reaction kinematics of both target nuclei are displayed in figure~\ref{bild.kinematik}
for the population of the 3/2$^-_1$-level at 646~keV and
the 1/2$^-_1$-level at 2.64~MeV in $^{37}$S.

The figure shows that in the $^2$H($^{36}$S,$^{37}$S$^*$)p reaction the
$^{37}$S nucleus is always
deflected by less than 2.5$^\circ$ and travels at about beam
velocity regardless of the deflection
angle. In contrast, the kinematics of the reaction with the
beryllium target shows a significantly larger deflection of $^{37}$S of up to 14$^\circ$
and a considerable energy variation. Hence one can expect
to achieve a sufficient Doppler correction when using a deuterium
target even without detecting one of the scattered reaction products, while at
least a partial kinematic
reconstruction is necessary to calculate the Doppler shift when a
beryllium target is used.

From conservation of
energy and momentum one finds that for a Q-value of
\begin{equation}
    Q>E_{\rm beam} \cdot \left(\frac{M_{\rm p}}{M_{\rm pt}}-1\right)
\end{equation}
the light reaction product can be emitted in backward direction.
Here $E_{\rm beam}$ is the energy of the beam particles and
$M_{\rm p}/M_{\rm pt}$ is the ratio of the projectile
masses before and after transfer. 
For the $^2$H($^{36}$S,$^{37}$S$^*$)p-reaction,
this condition is fulfilled for excited states up to an energy of 4.23~MeV, as the
reaction has a ground state Q-value of $+2.08$~MeV. 
For the $^9$Be($^{36}$S,$^{37}$S$^*$)$^8$Be reaction the condition is fulfilled for states up to
4.79~MeV, as the corresponding ground state Q-value is $+2.64$~MeV. 
Thus for both reactions and both excited states in
$^{37}$S considered, the emission of the light reaction product in backward
direction is possible, although its energy will be rather small.

\begin{figure}
\begin{center}
\resizebox{0.5\textwidth}{!}{%
\includegraphics{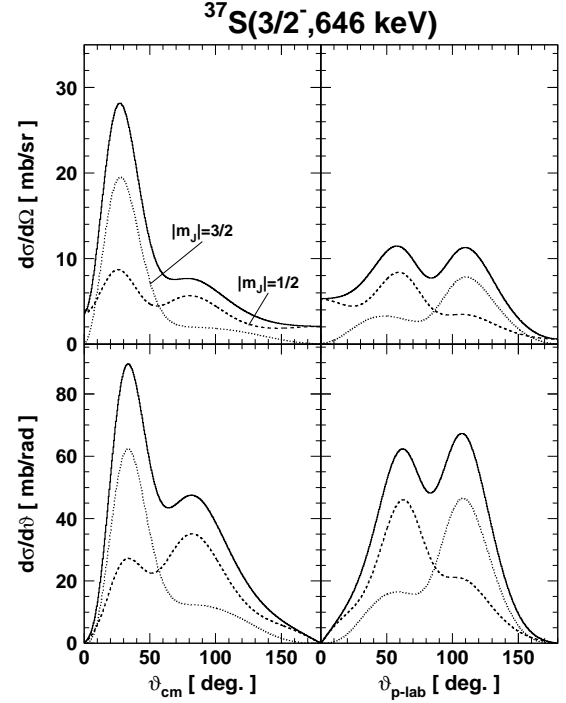}}
\end{center}
\caption[Differential Cross Sections of m-substates]
{Calculated differential cross sections for individual
   m-substates (dotted and dashed lines) and summed over the m-substates (solid line) of the
   3/2$^-_1$ level in $^{37}$S for the
   $^{2}$H$(^{36}$S$,^{37}$S$^*)$p-reaction at a beam energy of 2.2~MeV$\cdot A$. Left
   panels: differential cross sections in the c.m. system (with
   $\vartheta$ denoting the c.m. angle of $^{37}$S). Right panels:
   differential cross sections in the laboratory system when detecting the
   residual proton.
\label{bild.kinemat_wirkungsq_munter} }
\end{figure}

In the proposed transfer experiments the total cross sections to
excited $\gamma$-decaying states can be deduced from the
$\gamma$ lines integrated over the full solid angle covered by
MINIBALL.  
However, in order to determine the differential
c.m. cross sections (see figure \ref{bild.kinemat_wirkungsq_munter}) 
one has to trace one of the reaction products. 
If the variation of the laboratory angle and laboratory energy of the excited
transfer product with $\vartheta_\mathrm{cm}$ is sufficiently large, this can
be achieved by detecting its flight direction and using the
Doppler shift to distinguish between forward and backward scattering in
the c.m. system. 
An elegant alternative method is to analyze the
Doppler broadened $\gamma$ lineshapes observed around 0$^\circ$ or 180$^\circ$ 
with respect to the beam axis, which provide a direct measure of
the c.m. cross section modified by the particle-$\gamma$-correlation function
\cite{schwalm-pelte}. 
Both methods cannot be applied when using a $^2$H
target. 
Here it is only feasible to deduce information on the
differential cross sections by measuring at least the direction of the
light reaction product together with the decay $\gamma$ rays.

While the particle angular dependence of the differential transfer
  cross section
allows to determine the angular momentum of the transferred nucleon,
the total spin of the final states can be deduced from the
$\gamma$-angular distributions.
If necessary one can enhance the anisotropy of the $\gamma$ distribution by requiring
coincidences e.g. with all forward scattered particles.

\section{Setup of the experiment}
\begin{figure}
\begin{center}
\resizebox{0.45\textwidth}{!}{%
\includegraphics{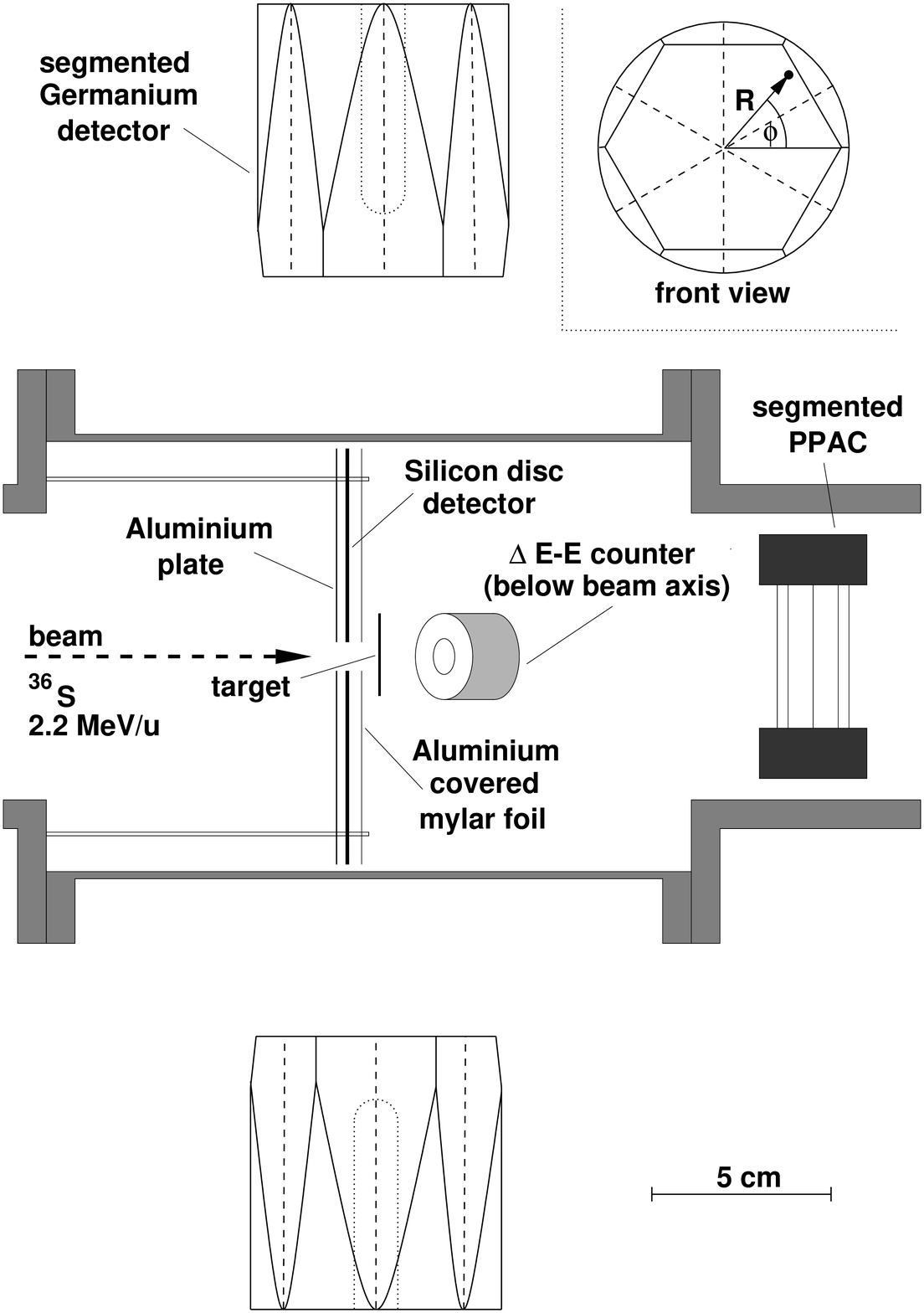}}
\end{center}
\caption{
   Experimental setup. 
   The tube-shaped target box is shown as cut in half. All detectors
   are seen from above. The  electric segmentation of the germanium
   detectors is indicated  by dashed lines. Detectors and target box are shown in correct
   proportions. 
   A front view of the germanium detector is shown in the inset and displays the definition of
   the radius $R$ and angle $\phi$ of the $\gamma$-ray interaction point.
\label{bild.aufbau}}
\end{figure}

The overall setup of the test experiment is shown in figure~\ref{bild.aufbau}.
Two six-fold segmented MINIBALL
germanium modules were used\footnote{For the measurement with the deuteron
target only one module was available.} and placed at an angle of
$\vartheta_{\rm det}=90^{\circ}$ with respect to the beam axis at a target
distance of 10.6~cm, which is
the minimum distance achievable with MINIBALL at REX-ISOLDE. 
It should be noted that not only the small detector-target distance but also 
the 90$^\circ$ detection angle correspond to the most demanding conditions
with regard to the Doppler shift correction of the recorded
$\gamma$ lines, as for this position the resulting $\gamma$-line width is most sensitive
to the finite granularity of the $\gamma$ detector.

The detector modules were equipped with preamplifiers for the six segment
signals and the core contact. 
The pulse heights of the charge signal at the core
preamplifier was used to obtain the $\gamma$-ray energy using
standard electronics and a peak sensing ADC. 
The energy information of the segments was obtained in the same way and used to deduce an information
about the $\phi$-angle of the $\gamma$-ray entry point (see inset in figure \ref{bild.aufbau}). 
Besides the energy information of the core and the segments, the
differentiated%
\footnote{A timing filter amplifier was used for differentiation.} 
preamplifier pulse of the core contact was recorded
using a 250~MHz 8-bit Flash-ADC and used to determine the radius $R$ of
the $\gamma$-ray entry point \cite{palafox} as defined in the inset in 
figure \ref{bild.aufbau}.
 
The REX-ISOLDE PPAC was positioned at zero degrees with the
entry foil 11~cm behind the target. 
The scattering angles covered by the PPAC reached up to
$\vartheta=9^{\circ}$,  so that all heavy reaction products of
the $^2$H($^{36}$S,$^{37}$S)p reaction passed through the detector 
as $\vartheta_{^{37}{\rm S}}<2.5^\circ$
(see figure~\ref{bild.kinematik}).  
In order to avoid damage to the PPAC a
mechanical shutter could be inserted in front of the PPAC when the
beam intensities exceeded the maximum allowed values of about $5\cdot10^7$ particles/s.

The PPAC consisted of 5 metallized
foils and was operated with isopropane gas at a pressure of 5~mbar.
The areal density ($<1.5$~mg/cm$^2$) was
sufficiently low to allow particles to traverse the detector without
stopping, so that the $\beta$ background and the $\gamma$ background
from $\beta$ decay was kept as low as possible.
At low beam intensities of up to around
10$^6$~particles/s it was possible to
count individual beam particles. 
At higher particle rates the detector
could still be used as a position-sensitive current monitor. A more detailed
description of this counter can be found in \cite{cub}.

In the single particle readout mode an efficiency of $>$99\% was measured
with $\alpha$-particles
(1.4~MeV$\cdot A$) from an $^{241}$Am source at a particle rate of
  10$^3$ s$^{-1}$.
In the in-beam experiment a decline of the efficiency was observed for
very high rates. 
At a particle rate of 10$^6$~particles/s
an efficiency of 70$\%$ (mainly due to deadtime problems)
was found. 
This efficiency was estimated by comparing the
peak area of the 646 keV $\gamma$ line from the decay of $^{37}$S
produced in the $^2$H ($^{36}$S, $^{37}$S$^*$)p reaction in a spectrum taken in
coincidence with the PPAC to a spectrum taken without this
coincidence. 
The efficiency decreased to 50$\%$ at a
rate of 1.5$\cdot 10^6$~particles/s, while the saturation rate was
3$\cdot 10^6$ particles/s.

In addition to the PPAC two auxiliary silicon particle detectors were employed.
A silicon ring detector with a thickness of 2~mm was placed
at a distance of 12~mm from
the target covering backward angles from $\vartheta=120^{\circ}$ to
$\vartheta=150^{\circ}$. The detector was used to investigate the
feasibility of observing the light reaction
products at backward directions. 
Although
it was possible to trace protons in backward direction, the low
particle energies and the
noise fluctuations of the silicon detector did not permit a
quantitative measurement of backscattered protons in this experiment. 

In forward direction a small $\Delta$E-E Si-detector-telescope with an active
area of 50~mm$^2$ and thicknesses of 50~$\mu$m and 1500~$\mu$m,
respectively, was located at a distance of 30~mm from the
target at an angle of $\vartheta=40^{\circ}$ with respect to the beam axis. 
The main
application of this detector was the selection of reaction channels by
identifying one of the light reaction products.
In the REX-ISOLDE setup this small telescope will be replaced by a large position-sensitive 
annular silicon disc-telescope detector \cite{edinbo}.
Both silicon detectors were covered with aluminized mylar foils with a
  thickness of 0.2~mg/cm$^2$. 

A $^{36}$S$^{8+}$ beam with an energy of 79.2~MeV (2.2~MeV$\cdot A$) was provided from the MPI-K tandem
accelerator. 
The beam was pulsed with a frequency of 13.56~MHz (74~ns) and a pulse width of $\sim$1~ns.
The beam intensity was varied between 2$\cdot 10^{5}$ and
6$\cdot 10^{8}$ particles per second.
Self-supporting targets of deuterium enriched polythene ($\sim$99\% CD$_2$, $\sim$0.5~mg/cm$^2$)
and $^9$Be (0.54~mg/cm$^2$) were used. 

\section{Feasibility Studies}
\subsection{Doppler Correction}\label{sec:dopp}
In MINIBALL the $\gamma$ detectors will be positioned close to the target to
optimize the $\gamma$-detection efficiency.   
As the recoil velocities encountered in inverse reactions are large, the Doppler broadening 
of the $\gamma$ lines, due to the large solid angle covered by a
single detector module, will therefore be severe.

In order to examine the capabilities offered by the segmented
germanium detectors in improving the $\gamma$-line widths,
the 646.2 keV line from the decay of the first excited state of $^{37}$S populated in the 
$^2$H($^{36}$S,$^{37}$S$^*$)p reaction at a beam energy of 2.2 MeV$\cdot A$ was
investigated in more detail. 
In the Doppler shift analysis, the
flight direction of 
$^{37}$S 
was assumed to be given by the beam axis and its 
average recoil velocity was taken to be $\beta=6.5\%$
of the velocity of light, thereby accounting for the energy loss of the S ions 
of 10\% in the target.

\begin{figure}
\begin{center}
\resizebox{0.5\textwidth}{!}{%
\includegraphics{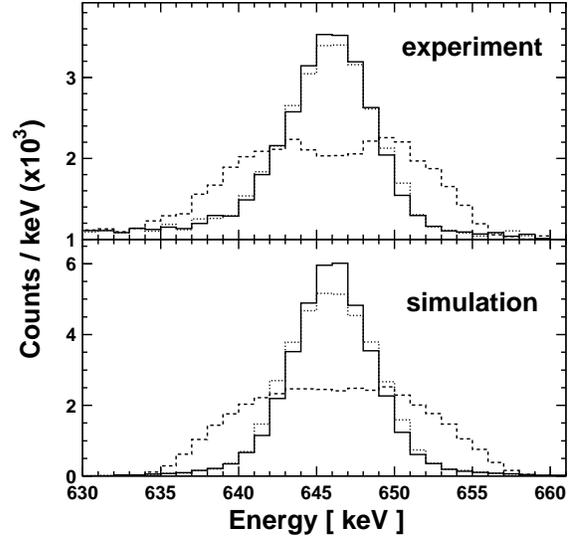}}
\end{center}
 \caption{Measured and simulated Doppler-corrected
   $\gamma$-lineshapes for the 646.2 keV transition in  $^{37}$S
  produced in the inverse reaction  $^2$H($^{36}$S,$^{37}$S)p. 
  The MINIBALL
  detector module was placed at $\vartheta_\mathrm{det}=90^{\circ}$ at 
  a distance of 10.6~cm. 
  (Dashed line: no entry point information used;
   dotted line: Doppler corrected $\gamma$ line using the segment information;
   solid line: Doppler corrected $\gamma$ line using both segment
  and radius information.)
  \label{bild.dopplerverglgemsim}}
\end{figure}
 
Without any additional information on the entry point of the $\gamma$ ray into
the detector an energy resolution of only
13.6~keV can be obtained for the  646.2~keV $^{37}$S line as compared to the intrinsic 
resolution of 1.7~keV
(see upper panel in figure~\ref{bild.dopplerverglgemsim}). 
Applying the $\phi$-angle algorithm (assuming an average
$\gamma$-ray entry radius of $R=20$~mm)
the $\gamma$-line width is reduced by a factor of 2 to 
6.7~keV. 
The angle algorithm only relies on the segment energy information and is based on the fact
that the first interaction point of the $\gamma$-ray in the detector
material, which carries the directional information,  
and the main interaction of the $\gamma$-ray, where the
maximum energy is deposited, are usually
located close together in the ($R$,$\phi$) space (see inset in figure \ref{bild.aufbau}). 
A full description of the algorithms used can be found in \cite{gund}.
If the 
additional radius information -- derived from the differentiated core
pulse shape
data with a resolution of $\pm$5~mm by applying the steepest-slope
algorithm \cite{palafox} --  is included
in the Doppler correction, the $\gamma$-line width can be
further decreased  to 6.2~keV.
The GEANT simulations shown in the lower panel of figure~\ref{bild.dopplerverglgemsim} 
give similar results (13.5~keV, 6.3~keV, and 5.6~keV, respectively). 
The small differences between the simulation and the measurement
remaining after considering also the intrinsic resolution of 1.7~keV
are due to the angle variation of the recoiling $^{37}$S nuclei of up to $\pm 2.5^{\circ}$,
which was disregarded in the simulation.

The final MINIBALL electronics will also support a detailed pulse shape analysis of the
segment signals. 
This allows for a more precise determination of the
position of the main interaction point of the $\gamma$ ray
\cite{gund,weisshaar} and therefore for a further improvement of the
Doppler corrected $\gamma$-line width.  
For the special case
considered here ($\vartheta_\mathrm{det}=90^{\circ}, \beta=0.065,
E_{\gamma}=646$~keV) and neglecting the intrinsic detector resolution of 1.7~keV,
we expect according to our simulations, an energy resolution
after Doppler correction of 3.6~keV if the flight direction of
the $\gamma$-emitting $^{37}$S nucleus is assumed to be exactly known ($0^\circ$).
A value of 4.7~keV is obtained when averaging over the recoil angles of the
$^{37}$S nuclei. 
It should be noted
again that the special case considered here corresponds to the most
demanding setup of a MINIBALL module with regard to the Doppler
correction. 
Considerably smaller line widths can be expected for all
those modules of the MINIBALL array positioned at angles 
$\vartheta_\mathrm{det} \not= 90^{\circ}$.

\subsection{Background Reduction}
\begin{figure}
\begin{center}
\resizebox{0.5\textwidth}{!}{%
\includegraphics{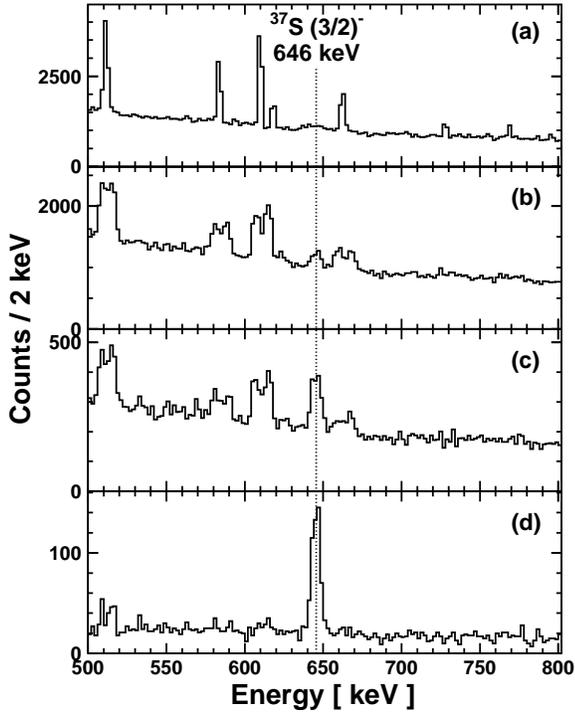}}
\end{center}
\caption{Partial $\gamma$-ray spectra following the $^{36}$S$\rightarrow$ $^2$H reaction.
The events were measured with one of the germanium
modules in a beam time of 7 hours at a beam intensity of 10$^6$
particles/s. 
In the panels is shown: 
(a) the raw $\gamma$-single spectrum; 
(b) the same as in panel (a), however, the Doppler correction relevant for the neutron pickup reaction was applied; 
(c) the same as (b), but a coincidence of the $\gamma$ signal with the accelerator RF was required; 
(d) the same as (c), but an additional coincidence with the PPAC signal was required, 
resulting in an essentially background free spectrum.
\label{bild.pplzvergl}}
\end{figure}

The background suppression achieved with the REX-ISOLDE PPAC 
was investigated using the $^2$H($^{36}$S,$^{37}$S$^*$)p reaction
at low beam intensities of 10$^6$~particles/s.
The results are 
displayed in figure~\ref{bild.pplzvergl}. 
In the raw germanium spectrum, shown in panel (a), the 646 keV deexcitation
$\gamma$ line of the first excited state in $^{37}$S is hardly visible
between the narrow background peaks.
When the Doppler correction is applied (panel (b)) the 646 keV line
becomes more pronounced as
its width is  reduced, while
the background lines are broadened. 
The deexcitation line is now visible on top of the background with a
peak-to-background ratio of 10\%. 
Requiring a coincidence of the germanium signal with the RF signal, controlling the
beam pulsing system of the accelerator, the background is
reduced by more than a factor of 5 (panel (c)) as the Ge-detector is
no longer active in between the beam pulses. 
The peak areas
of the 646 keV line in both spectra agree within statistical limits, i.e. the detector efficiency is
not reduced by the coincidence requirement.

An almost background free $\gamma$-spectrum is obtained when an additional coincidence
with the PPAC is required (panel (d)). 
Although the peak area of the 646 keV line is suffering a loss of about 30\% due to the 
reduced efficiency of the PPAC at particle rates of $10^6$~particles/s,
the background is reduced by another factor of $\sim13$. 
This reduction is caused by the fact that for a beam-pulse frequency of 13.56~MHz
and a beam intensity of 10$^6$~particles/s, on the average only every 14th beam
pulse contains a projectile. 
Thus by requiring the PPAC coincidence, background $\gamma$ rays accumulated when an empty beam
bunch arrives are rejected. 
Note that no background spectra have been subtracted from the spectra shown in  
figure~\ref{bild.pplzvergl}. 
By doing so, the small
remaining background in figure~\ref{bild.pplzvergl}d can be quantitatively
removed.

\section{Results}
\subsection{Reaction Channels on the CD$_\mathbf 2$ Target}
Bombarding a deuterium target with beams having an energy of 2.2 MeV$\cdot A$ $^{36}$S,  
the two strongest reaction channels will be the neutron and the proton
pickup.
As REX-ISOLDE is designed to explore new regions of unknown nuclei, this
is an important advantage, as for the assignment of the observed
$\gamma$ transitions only two possible nuclei have to be considered. 
Furthermore, the
resulting transfer nuclei travel at about the same speed and in the same
direction as the beam particles allowing a Doppler
correction of the $\gamma$ lines without the need to determine the
kinematic by particle detection.
However, pure deuteron targets are not readily available.
Thus deuterium enriched polythene
(CD$_2$)$_\mathrm{n}$ is widely used as a target material. Polythene
is easy to handle and foils of any desired thickness can be
produced.
However, additional reaction channels may occur due to the
presence of the carbon.

\begin{figure}
\begin{center}
\resizebox{0.5\textwidth}{!}{%
\includegraphics{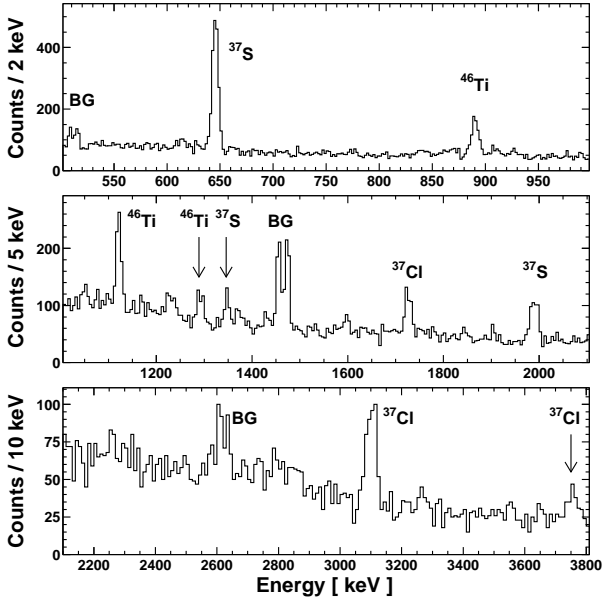}}
\end{center}
\caption{Doppler-corrected PPAC-gated germanium spectrum
without background subtraction observed in the 
$^{36}$S$\longrightarrow$(CD$_2$)$_n$ reaction at 2.2~MeV$\cdot A$. 
The spectrum was taken in a 24 hours beamtime at a beam intensity of 1$\cdot10^6$~particles/s
with a single germanium detector at $\vartheta_{\rm
det}=90^\circ$.  The spectrum corresponds to that of
figure~\ref{bild.pplzvergl}d, however, the full energy range and the
full statistics are shown.
\label{bild.detpplz_s12-20}
}
\end{figure}
 
In figure~\ref{bild.detpplz_s12-20} the Doppler-corrected
$\gamma$ spectrum observed at $\vartheta_\mathrm{det}=90^{\circ}$ in 
coincidence with the PPAC and the accelerator
RF when bombarding a (CD$_{2}$)$_n$ target with a $^{36}$S beam is
displayed. 
The spectrum was recorded in 24 hours at a beam intensity of
1$\cdot10^6$~particles/s and a beam energy of
2.2~MeV$\cdot A$.
The Doppler correction was performed as discussed in section \ref{sec:dopp}.

Prominent lines from the deexcitation of the neutron pickup product $^{37}$S 
occur at 646~keV, 1347~keV, and 1993~keV. At 1727~keV and at 3117 keV
 $\gamma$ lines from the proton pickup product $^{37}$Cl are observed, the
 latter line being composed of two known $^{37}$Cl lines at
3086~keV and 3103~keV. 
The lines at 890~keV, 1122~keV, and 1289~keV
are due to $\gamma$-transitions in $^{46}$Ti
populated in the fusion-evaporation reaction
$^{12}$C($^{36}$S,2n)$^{46}$Ti. 
A list of all strong $\gamma$ transitions observed,
their assignment, and their relative $\gamma$ intensities 
is given in table~\ref{tabelle.cd2population}.

In the reaction investigated in this test experiment all
reaction products are well known, however,
when exploring unknown regions of the nuclear chart it is necessary
to distinguish the $\gamma$-transitions emitted from the transfer
products from
background and $\gamma$-transitions caused by other reaction channels.

Background $\gamma$-lines are easily recognized as these
lines become broader after the Doppler correction. Thus they can be immediately
identified and eliminated by background-subtraction.
The background spectrum can be obtained by recording a
$\gamma$-single spectrum in anti-coincidence with the accelerator RF
and the PPAC.
 
The different Doppler shifts of $\gamma$-lines emitted from transfer
and fusion products can also be used to distinguish between these two channels. 
When $\gamma$-lines produced in fusion reactions are Doppler
corrected with the transfer kinematics the resulting peak position
will still depend on the $\gamma$-observation direction. 
As a result the fusion lines in the Doppler-corrected sum spectra obtained 
e.g. by summing over all MINIBALL modules in the forward and in the backward
direction separately will be broadened and centered at different energies.

It is more demanding to separate the two pickup channels 
since, unless the Q-values are very different (as is the case for
  exotic nuclei, but not here), 
they show the same kinematics and Doppler behavior.
If a $\gamma$-transition is already known in one of the nuclei $\gamma$-$\gamma$-coincidences may help
to identify further transitions. 
Otherwise, one can either use the
large annular silicon detector available in REX-ISOLDE experiments in
the forward hemisphere to detect the proton leftover in the neutron pickup
reaction, or perform the same measurement with the $^{9}$Be target,
where the proton pickup is suppressed (see section \ref{kap.beryllium}).

\subsection{Cross Section Determination for the CD$_2$ Target}
The experimental setup allows to determine absolute cross
sections rather easily, since the total number of beam particles which have passed the
target during beamtime is available from the REX-ISOLDE PPAC. 
Together with the number of counts in a $\gamma$ line at energy E$_{\gamma}$ (deduced from the
$\gamma$ spectrum observed in coincidence with the PPAC), the
known areal density of the deuterium in the CD$_2$ target,  and
the efficiency of the germanium detectors $\varepsilon_{\rm Ge}$(E$_{\gamma}$) it
is possible to derive the absolute cross
sections $\sigma_{\gamma}$(E$_{\gamma}$) for the $\gamma$ line by:
\begin{eqnarray} \label{wq.ppac}
\sigma_{\gamma}(E_\gamma) & = &
\frac{N_{\gamma -\rm PPAC}(E_\gamma)/ (\varepsilon_{\rm Ge}(E_\gamma)\cdot\varepsilon_{\rm PPAC})}
     {p\cdot(N_{\rm PPAC}/ \varepsilon_{\rm PPAC})\cdot (N/A)_{\rm deuteron}} \nonumber \\
& = & \frac{N_{\gamma -\rm PPAC}(E_\gamma)/\varepsilon_{\rm Ge}(E_\gamma)}
              {p\cdot N_{\rm PPAC}\cdot (N/A)_{\rm deuteron}} \; .
\end{eqnarray}
Here $N_{\gamma-\mathrm{PPAC}}(E_\gamma)$ is the number of $\gamma$ rays, which
were observed in the germanium detector in coincidence with the PPAC,
$\varepsilon_{\rm Ge}(E_\gamma)$ the energy-dependent full-energy peak efficiency of the
germanium detector, $\varepsilon_{\rm PPAC}$ the detection
efficiency of the PPAC, $N_{\rm PPAC}$ the number of beam
particles counted by the PPAC, and $(N/A)_{\rm deuteron}$ denotes the deuteron areal density in
the target given in atoms/cm$^2$. 
The areal density of the CD$_2$
target of 0.55~mg/cm$^2$, used in the present experiment, corresponds to $(N/A)_{\rm
deuteron}=4.1\cdot 10^{19}{\rm cm}^{-2}$ and $(N/A)_{^{12}{\rm C}}=2.1\cdot 10^{19}{\rm cm}^{-2}$.
The factor $p$ takes into account that due to the time structure of the beam
(1~ns wide pulses every 74~ns) there exist a non-zero probability that
more than one projectile is contained in one beam pulse, which cannot
be distinguished by the Ge-RF-PPAC coincidence requirement.
In the present experiment $p$ is approximately given by $p=1.07(2)$.
 
It is important to note that the efficiency of the PPAC cancels out so
that this method allows to derive absolute cross sections without
the knowledge of the efficiency of the PPAC. 
One should also be aware that in equation~(\ref{wq.ppac}) a
possible angular distribution of the emitted $\gamma$-rays with 
respect to the beam direction is
neglected; the formula is exact for isotropic $\gamma$-ray
emission, otherwise  a correction factor is
needed. 
However, this factor reduces to 1 if the 
full MINIBALL array is used as the array covers all
regions of the full 4$\pi$ solid angle equally well.

Using equation~(\ref{wq.ppac}) and the intensity data accumulated in the 24~hour beamtime 
at a beam intensity of 10$^6$~particles/s,
an absolute cross section of
215(43)~mb has been derived for the 646~keV
$\gamma$ line of $^{37}$S assuming an isotropic $\gamma$-angular
distribution (see further below).
The uncertainty of the cross section value comprises statistical errors, the error of
the target thickness (5\%), and the uncertainty in the efficiency 
of the germanium detector (15\%).

\begin{table*}
\begin{center}
\caption{Compilation of the strongest $\gamma$ lines observed at $\vartheta_\mathrm{det} = 90^\circ$ 
in the reaction $^{36}$S$\rightarrow$CD$_2$ at an energy of 2.2 MeV$\cdot A$.
\label{tabelle.cd2population}}
\begin{tabular}{|rr|c||c|c|c|}
\hline
\multicolumn{2}{|c|}{E$_\gamma$ $[{\rm keV}]$} & N$_{\gamma\rm}$ & reaction \& transition &
$\frac{\sigma_{\gamma} ({\rm E})}{\sigma_{\gamma} (646)}$ &
$\sigma_{\gamma} ({\rm E_\gamma}) [{\rm mb}]$ \\ \hline\hline
646(1)  & ({\sl 646}) & 19500 (5$\%$)     & $^{37}$S : $3/2^-_1\to7/2^-_{\rm gs}$ & 100\%    & 215(43) \\
 ---    & ({\sl 751}) & $<$500            & $^{37}$S : $3/2^+_1\to3/2^-_1$        & $<$3$\%$ & $<$5 \\
1347(3) & ({\sl 1346}) & 1000 (40$\%$)    & $^{37}$S : $3/2^-_2\to3/2^-_1$        & 8$\%$    & 19(10) \\
 ---    & ({\sl 1377}) & $<$400           & $^{37}$S : $7/2^-_1\to3/2^-_1$        & $<$3$\%$ & $<$5 \\
        & ({\sl 1992}) &                  & $^{37}$S : $3/2^-_2\to7/2^-_{\rm gs}$ & &  \\
\raisebox{1.5ex}[1.5ex]{1993(2)$^\dagger$} &\raisebox{1.5ex}[1.5ex]{$\Big\{$} ({\sl 1992})  & \raisebox{1.5ex}[1.5ex]{3100 (10$\%$)} 
                                          & $^{37}$S : $1/2^-_1\to3/2^-_1$ 
                                 &\raisebox{1.5ex}[1.5ex]{ 29$\%$}       &\raisebox{1.5ex}[1.5ex]{61(14)} \\
---& ({\sl 2023}) & $<$300                & $^{37}$S : $7/2^-_1\to7/2^-_{\rm gs}$  & $<$3$\%$ & $<$5 \\ \hline
1727(3) & ({\sl 1727}) & 1800 (20$\%$)    & $^{37}$Cl : $1/2^+_1\to3/2^+_{\rm gs}$ & 16$\%$   & 33(10) \\
&({\sl 3086}) &                           & $^{37}$Cl : $5/2^+_1\to3/2^+_{\rm gs}$ & &  \\
\raisebox{1.5ex}[1.5ex]{3101(5)$^\dagger$} &\raisebox{1.5ex}[1.5ex]{$\Big\{$} ({\sl 3103}) & \raisebox{1.5ex}[1.5ex]{3200 (10$\%$)} 
                                          & $^{37}$Cl : $7/2^-_1\to3/2^+_{\rm gs}$ &
                            \raisebox{1.5ex}[1.5ex]{45$\%$} &\raisebox{1.5ex}[1.5ex]{98(23)} \\
3748(10)& ({\sl 3741}) & 300 (50$\%$)     & $^{37}$Cl : $5/2^-_1\to3/2^+_{\rm gs}$ & 5$\%$ & 9(5) \\ \hline
 890(1) & ({\sl 889}) & 6800 (8$\%$)      & $^{46}$Ti : $2^+_1\to0^+_{\rm gs}$ & 85$\%$    & 187(42) \\
1049(3) & ({\sl 1049}) & 800 (40$\%$)     & $^{46}$Ti : $3^-_1\to4^+_1$        & 11$\%$    & 23(9) \\
1122(1) & ({\sl 1121}) & 5500 (10$\%$)    & $^{46}$Ti : $4^+_1\to2^+_1$        & 77$\%$    & 168(37) \\
1289(2) & ({\sl 1289}) & 1900 (20$\%$)    & $^{46}$Ti : $6^+_1\to4^+_1$        & 28$\%$    & 61(19) \\\hline
\multicolumn{6}{l}{$\dagger$ the two transitions could not be resolved}
\end{tabular}
\end{center}
In the different columns are listed:\\
1) the transition energies extracted from Doppler-corrected spectra using the reaction-specific 
kinematics;  literature values are given in italics;\\
2) the number of $\gamma$ rays observed in a high statistics run performed without PPAC coincidences;\\
3) the deduced transition, based on the level energy: $^{37}$S: $^2$H($^{36}$S,$^{37}$S$^*$)p,
  $^{37}$Cl: $^2$H($^{36}$S,$^{37}$Cl$^*$)n, $^{46}$Ti: $^{12}$C($^{36}$S,$^{46}$Ti$^*$2n);\\
4) the relative cross sections assuming isotropic emission of the
  $\gamma$ rays;\\
5) absolute cross sections (see text).\\
\end{table*}

With the absolute cross section for the 646 keV line the
cross sections can be deduced for other $\gamma$ lines from the
  relative $\gamma$ intensities,
which were deduced in the present case from a high statistic run without requiring
PPAC-coincidences (see table~\ref{tabelle.cd2population}).  

While the cross sections $\sigma_\gamma(E_\gamma)$ for the occurrence of a
$\gamma$ ray of energy $E_\gamma$ are directly
accessible,
the relevant spectroscopic data is the
cross section $\sigma(E)$ to directly populate a certain
excited level of energy $E$. 
Both kinds of cross sections are linked by the branching ratios for
the different deexcitation channels. 
While the branching ratios 
for the decay of excited states of $^{37}$S populated in this experiment are
already known, information about the level scheme and the branching ratios
of unknown nuclei investigated in REX-ISOLDE experiments will have to be
extracted from the measurement.

\begin{figure}
\begin{center}
\resizebox{0.45\textwidth}{!}{%
\includegraphics{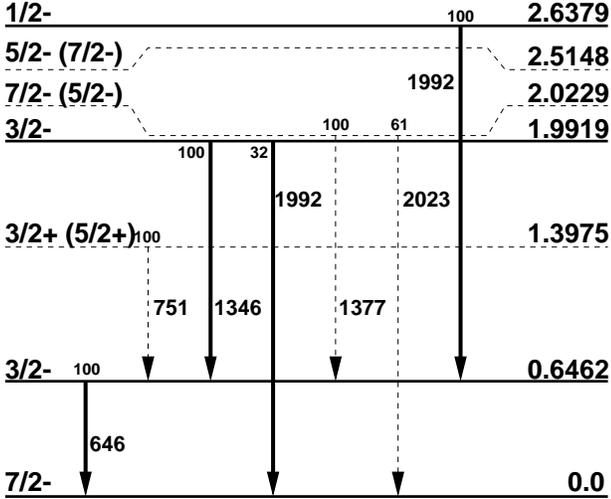}}
\end{center}
\caption{Level scheme of $^{37}$S. The $\gamma$ transitions observed are marked
  by thick arrows, levels populated are shown with bold lines. 
Level energies and branching ratios are taken from \cite{fir96}.
  \label{bild.s37level}}
\end{figure}
 
With the aid of 
the level scheme of $^{37}$S and the branching ratios
(see figure~\ref{bild.s37level}) 
the cross section $\sigma(E)$ was calculated.
The population of three negative parity low spin levels was
observed:
the absolute cross section for the population of the 3/2$^-_1$-level
  at 646~keV, the 3/2$^-_2$-level at 1992~keV, and the 1/2$^-_1$ level at 2638~keV 
was 140(46)~mb, 23(10)~mb, and 56(15)~mb, respectively 
(see table~\ref{tabelle.wirkungsquerschnitte}).
The two states
which were most strongly populated correspond to the
$^{36}$S core and an
excited neutron in the 2p$_\frac{3}{2}$ (3/2$^-_1$) and
  2p$_\frac{1}{2}$ (1/2$^-_1$) shell.
A simple classical estimate, which assumes a $^{36}$S-nucleus, a
single neutron, and a relative velocity corresponding to an energy of
2.2~MeV$\cdot A$, shows that the most probable angular momentum transfer is
1$\hbar$.
Hence, transfer to excited states with high orbital
angular momenta are expected to be suppressed; as a consequence
the population of the 1f$_{5/2}$ and 1f$_{7/2}$ single neutron states
was not observed.

\begin{table}
\begin{center}
\begin{tabular}{|c|c||c|c|c|}
\hline
E$_\mathrm{level}$  & J$^{\pi}$ & $\sigma_{\rm exp.}$  & $\sigma_{\rm theo.1}$  &
$\sigma_{\rm theo.2}$  \\
$[{\rm keV}]$&&$[{\rm mb}]$&$[{\rm mb}]$&$[{\rm mb}]$\\
\hline
646 & $3/2^{-}$ & 140(46) & 180 & 105 \\
1992 & $3/2^{-}$ & 23(10) & - & - \\
2638 & $1/2^{-}$ & 56(15) & 95 & 75 \\
\hline
\end{tabular}
\end{center}
\caption{Comparison of the experimental and calculated total cross sections
for three excited states in $^{37}$S populated in $^{2}$H$(^{36}$S$,^{37}$S$^*)$p 
neutron pickup reaction at 2.2 MeV$\cdot A$.
\label{tabelle.wirkungsquerschnitte}}
\end{table}

The measured cross sections were compared with theoretical
predictions (listed in table \ref{tabelle.wirkungsquerschnitte}). 
Skyrme-Hartree-Fock calculations were used to derive the structure of the nuclei. 
The results were used to obtain transfer probabilities using
the Distorted-Wave-Born-Approximation (DWBA)
and the Exact-Finite-Range-Distorted-Wave-Born-Approximation
(EFR-DWBA)(for details see \cite{lenske}).  
When the level energies obtained in the theoretical calculations were
used, theoretical model cross sections of $\sigma(3/2^-_1)=180$~mb and
$\sigma(1/2^-)=95$~mb were derived, whereas cross sections of 
$\sigma(3/2^-_1)=105$~mb and $\sigma(1/2^-)=75$~mb were found
when the level energies were fixed at their experimental  values.
The agreement of the latter results with the experimental values is fair.

For the calculations of the experimental cross sections 
isotropic angular distributions of the $\gamma$ rays emitted from the 
excited nuclei were assumed. 
While this assumption is always true for $J=1/2$ states, it applies  
for the 3/2 levels only, if all m-substates are populated equally.  
Although the differential cross sections to populate the $|$m$|=1/2$  and 
the $|$m$|=3/2$ substates of the 3/2$^-_1$ level differ
significantly, as can be seen in
figure~\ref{bild.kinemat_wirkungsq_munter}, the angle integrated total
probabilities to populate these substates
are 48~\% and 52~\% respectively, and the angular distribution of the
emitted $\gamma$ rays deviates by less than 1\% from an isotropic
distribution. 
Even if one assumed complete alignment, the deviation
would be less than 14\%.

In the anticipated REX-ISOLDE experiments the full MINIBALL will be used,
which subtends the 4$\pi$ solid angle of the deexcitation $\gamma$ rays rather isotropically.
This will facilitate the determination of the absolute cross sections $\sigma_\gamma(E_\gamma)$,
the branching ratios needed to derive the corresponding total cross section $\sigma(E)$
as well as the differential cross sections $d\sigma/d\Omega$.
Moreover, $\gamma$ angular distributions, $\gamma$-$\gamma$ and $\gamma$-particle 
correlations are in principle accessible allowing to derive level schemes, spins, and parities.
In favorable cases even lifetime measurements will be possible (see e.g. \cite{schrieder}).

\subsection{Beryllium Induced Transfer}
\label{kap.beryllium}
In contrast to the deuterium target the proton pickup is
strongly suppressed as compared to the neutron pickup when using the $^{9}$Be target, thus
facilitating the identification of the latter reaction channel. 
Yet $^9$Be exhibits two significant
disadvantages as a target.
Besides the neutron pickup reaction $^9$Be($^{36}$S,$^{37}$S$^*$)$^8$Be 
there exist several fusion-evaporation channels.
Furthermore $^9$Be is significantly heavier than deuterium
and the reaction products are deflected by larger angles and are
subject to larger energy variations (see figure~\ref{bild.kinematik}), 
and a good Doppler correction can only be achieved by tracking the direction of one of
the reaction products. 
For the present test measurement an 8 hour run with a high intensity (6$\cdot10^8$~particles/s)
$^{36}$S beam  of 2.2~MeV$\cdot A$ was performed, because in the
present setup only
a small silicon telescope detector was available for this task. 
Consequently, the
REX-ISOLDE PPAC had to be protected against the beam particles with an aluminum shutter.

\begin{figure}
\begin{center}
\vspace{-1.cm}
\resizebox{0.5\textwidth}{!}{%
\includegraphics{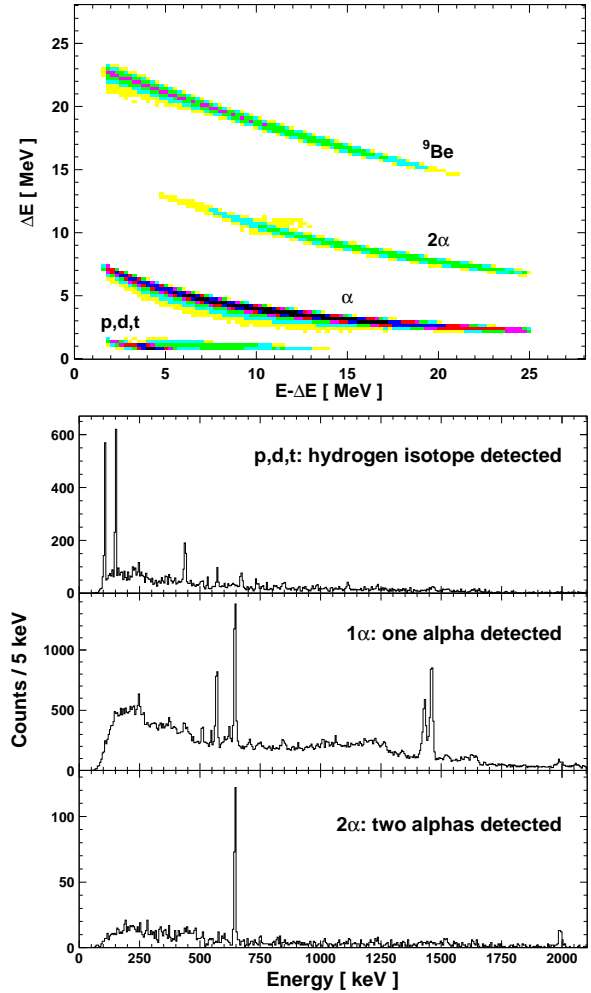}}
\end{center}
\caption{Doppler corrected $\gamma$ spectra observed in the
$^{36}$S$\longrightarrow^9$Be reaction at 2.2~MeV$\cdot A$ are shown in the bottom plot.
The spectra were generated by gating on the particle branches identified 
in the $\Delta E$--($E-\Delta E$) matrix of the silicon telescope detector (top plot).
\label{bild.beryllium_allin1}
}
\end{figure}

As manifested in figure~\ref{bild.beryllium_allin1}, the silicon $\Delta$E-E
detector is a major aid to distinguish between different reaction
channels and to perform the proper Doppler shift correction. 
Putting a gate on the hydrogen isotopes yielded several $\gamma$ transitions
mainly at low energies, which could be assigned to potassium isotopes corresponding to the
$^9$Be($^{36}$S,p2n)$^{42}$K and
$^9$Be($^{36}$S,pn)$^{43}$K reactions (top panel in bottom plot of figure~\ref{bild.beryllium_allin1}).
Gateing on one $\alpha$-particle $\gamma$-rays of argon isotopes produced in the
$^9$Be($^{36}$S,$\alpha$2n)$^{39}$Ar and the 
$^9$Be($^{36}$S,$\alpha$n)$^{40}$Ar reaction
as well as of $^{37}$S from the neutron pickup reaction are observed. 
Requiring
both $\alpha$ particles originating from the $^8$Be decay
the $^{37}$S neutron pickup channel
$^9$Be($^{36}$S,$^{37}$S$^*$)2$\alpha$ can be selected exclusively. 
The intensity drop of the 646 keV $^{37}$S line when requiring the detection
of both $\alpha$ particles as compared to one $\alpha$ particle is due to 
the small solid angle of the telescope detector used.
In the REX-ISOLDE setup this efficiency loss will be substantially smaller
because of the large solid angle subtended by the annular telescope 
detector planned to be employed in these measurements.
The $\gamma$ transitions not shown in
the gated spectra could be assigned to calcium
isotopes from the fusion-evaporation channels
$^9$Be($^{36}$S,3n)$^{42}$Ca and
$^9$Be($^{36}$S,2n)$^{43}$Ca.
An overview of the strongest $\gamma$ transitions detected is given in \cite{gund}.

It is interesting to note
that the $^{43}$K line at 738~keV has a long lifetime of 200~ns. As
the beam particles travel about 2~cm/ns and the
PPAC was covered with an aluminum shutter plate in this measurement, the line
is only visible as a sharp peak if no Doppler correction is performed
  as most nuclei $\gamma$ decay after being stopped in the shutter. 
For REX-ISOLDE the population of
excited states with a lifetime of more than a few nanoseconds will pose
problems, as the reaction products will not be stopped in the target
chamber and their  decay $\gamma$ rays will therefore remain undetected.

As no low intensity
measurements with the PPAC were performed using the Beryllium target, absolute cross sections for the
$^9$Be($^{36}$S,$^{37}$S$^*$)$^8$Be reaction could not be deduced,
but the relative $\gamma$ intensities
for the different nuclei and states populated in the
$^{36}$S$\longrightarrow^9$Be reaction could be determined.
The fusion-evaporation is the strongest channel leading mainly
to $^{42}$Ca (85\%) and $^{43}$Ca (40\%), $^{42}$K (50\%) and $^{43}$K (25\%) as well as
 $^{40}$Ar (100\%) and  $^{39}$Ar (5\%). The
neutron pickup leading to $^{37}$S (100\%) served as the reference cross section.
As expected, no $^{37}$Cl
lines from the proton pickup channel were observed;
for the population probability an upper limit of 2\% can be given.  

\section{Summary}
A test experiment was performed to investigate the feasibility of $\gamma$-ray
spectroscopy experiments using inverse transfer
reaction with low-intensity radioactive beams.
 
Two single modules of the MINIBALL germanium array were used to
detect the deexcitation $\gamma$ rays.
Exploiting the six-fold segmentation of the detector modules a relative $\gamma$-energy
resolution of $<1\%$ could be achieved after the Doppler shift correction, even though
the recoil velocities of the $\gamma$-emitting nuclei were
as large as $\beta=0.065$ and the detectors were placed at
a distance of only 10.6~cm from the target at the most unfavorable detection angle of $90^\circ$ 
with respect to the recoiling nuclei.
This is an improvement of the $\gamma$-line width by
more than a factor of 2 as compared to a module without segmentation and
when no pulse shape analysis is used. 
A further improvement of the $\gamma$-energy resolution by almost another factor of 2 will be possible when
the new MINIBALL electronics will be available. 
The new electronics will allow to exploit the signal shapes of the segment pulses 
(in addition to the core pulses) to further improve the $\gamma$-ray entry point determination.

The new REX-ISOLDE PPAC delivered a time reference signal for
particle-$\gamma$-coincidence measurements allowing to perform 
$\gamma$ spectroscopy at very low rates. 
The $\gamma$ rays detected
without a beam particle were rejected, so that
the peak-to-background ratio
became independent of the beam particle rate.
Furthermore, the PPAC allowed to
count individual beam particles up to a rate of 
10$^6$~particles/s.
Hence an easy method to determine absolute cross sections was
available, as the efficiency of the PPAC does not need to be known.

In the $^{36}$S$\longrightarrow$(CD$_2$)$_n$ reaction three reaction
channels were observed. Besides the deuteron induced neutron and proton
transfer channels leading to $^{37}$S and $^{37}$Cl also the
fusion-evaporation channel $^{12}$C($^{36}$S,2n)$^{46}$Ti is found
to contribute to the $\gamma$ spectrum. 
Due to the different Doppler shift of the $\gamma$ lines it is
possible to distinguish between fusion and pickup reactions.
The additional identification of the neutron pickup channel requires either 
the coincident detection of the proton or an additional 
measurement with a $^9$Be target.

Model calculations to predict the neutron pickup 
cross sections were performed for two states populated in the 
$^2$H($^{36}$S,$^{37}$S$^*$)p reaction. 
The results agree reasonably well with the measurements. 

The $^9$Be target gives complementary information about the
neutron pickup reaction. 
In addition, neutron-rich compound nuclei are
populated, which will be of considerable interest as well when using
neutron-rich exotic beams.
However, it is mandatory in this case to have a position-sensitive silicon telescope available to
detect the light reaction products to be able to distinguish between the
different reaction channels and to perform the proper Doppler shift correction.

The present experiment has clearly demonstrated that
it is possible to investigate exotic nuclei at
REX-ISOLDE using pickup reactions and $\gamma$-ray spectroscopy.
When MINIBALL is fully equipped and a CD$_2$ target with an areal density of
0.5~mg/cm$^2$ is used, it will take 20 hours to accumulate 400
counts in a $\gamma$ line at 1~MeV assuming a beam intensity of 10$^4$
particles/s, a cross section of $\sigma_\gamma(E_\gamma)=100$~mb and a
MINIBALL efficiency of 14\%.

This work was partly supported by the BMBF under contract No 06DA915I and the 
DFG under contract No Le439/4.

\end{document}